# Self-organization into quantized eigenstates of a classical wave driven particle

Stéphane Perrard[1], Matthieu Labousse[1], Marc Miskin[1,2,3], Emmanuel Fort[2,*], and Yves Couder[1]

[1]*Matières et Systèmes Complexes, Université Paris Diderot, CNRS UMR 7057, 10 rue Alice Domon et Léonie Duquet, 75205 Paris Cedex 13, France, EU*

[2]*Institut Langevin, ESPCI ParisTech, CNRS UMR 7587, 1 rue Jussieu, 75238 Paris Cedex 05, France, EU*

[3]*Current address: James Franck Institute, University of Chicago, 929 East 57th Street, Chicago, IL 60637*

*Corresponding author: emmanuel.fort@espci.fr*

**A growing number of dynamical situations involve the coupling of particles or singularities with physical waves. In principle these situations are very far from the wave-particle duality at quantum scale where the wave is probabilistic by nature. Yet some dual characteristics were observed in a system where a macroscopic droplet is guided by a pilot-wave it generates. Here we investigate the behaviour of these entities when confined in a two-dimensional harmonic potential well. A discrete set of stable orbits is observed, in the shape of successive generalized Cassinian-like curves (circles, ovals, lemniscates, trefoils...). Along these specific trajectories, the droplet motion is characterized by a double quantization of the orbit spatial extent and of the angular momentum. We show that these trajectories are intertwined with the dynamical build-up of central wave-field modes.  These dual self-organized modes form a basis of eigenstates on which more complex motions are naturally decomposed.**

There is a large variety of physical systems in which localized objects (particles, point-like sources or singularities) are associated with extended waves. These entities have a generic character being observed in *e.g.* localized soliton-like dissipative structures, defects in extended non-linear patterns, polaritons, electromagnetic knots, chemical waves or in the vortices of Bose Einstein condensates[1-9]. Their self-organization is of particular interest because of the specific non-locality of their wave-mediated interactions. Our group has introduced a simple experimental system having such an interplay between a wave and a particle endowed with an additional memory effect. The particle is a liquid drop, bouncing on a vertically vibrating bath. It is propelled by its interaction with the surface waves that it produces[10,11]. The resulting dynamical entity is designated henceforth a walker. It is interesting to note that the walker implements a pilot-wave system. This idea was first proposed theoretically by de Broglie in the early stage of quantum physics, as an attempt to predict the wave-like properties of elementary particles[12,13]. For de Broglie, the localized object (particle or singularity) moved, guided by a spatially extended pilot-wave. In de Broglie's model the pilot wave was a real wave; later Bohm[14] suggested that it could rather be the solution of the Schrödinger equation. These models, intrinsically different from each other[15,16], remain two important hypotheses for a possible interpretation of quantum effects.

The walker pilot-wave has two assets; it is a real wave and it is associated with a memory. As the droplet bounces on a vibrated bath, each impact generates a packet of parametrically forced Faraday waves that is sustained for a finite time[11]. The total wave-field that guides the droplet is a linear superposition of these waves. Its interference pattern contains a memory of the past trajectory. Previous experiments on walkers have revealed that when the dynamics is dominated by memory effects, some quantum-like properties arise. They include independently a form of uncertainty principle[17,18] and a form of discretization[19,20]. The latter was obtained in an experiment in which a walker moving on a rotating bath is submitted to a Coriolis force. At long memory the motion is affected by the associated waves: the orbit remains circular but its radius can only take discrete values. In this situation, due to the translational invariance, only one integer is needed to completely define the observed states. Assuming in this experiment that the Faraday wavelength played the role of the de Broglie wavelength in quantum mechanics, this discretization was shown to have a similarity with angular momentum quantization[18].

These results are surprising regarding the huge differences between walkers and quantum particles. Walkers have a classical scale so that they have no relation to the Planck constant. They are not Hamiltonian: they are dissipative structures sustained by an external forcing. Furthermore, their associated waves are not probability waves but physical waves propagating on a material substrate. Precisely because of this distance, the previous results obtained with walkers suggest that some phenomena could be general enough to overcome the classical-quantum barrier and be common to the two worlds. If this can be proven true, experiments performed at classical scale could shed light on non-resolved quantum phenomena. Furthermore, in this case contrary to Landau quantization, there are two degrees of freedom that cannot be reduced. Correspondingly, we find that two integers are needed to fully characterize a single state. This double quantization opens the possibility of degeneracy, which is observed experimentally. Here we thus demonstrate the existence of a classical analogue of quantum eigenstates that was still missing. The results obtained hitherto suggested that the most important asset of a walker is its interaction with its own past whenever it visits a region it has already disturbed. This type of spatio-temporal non-locality opens the possibility of self-organization for a single object. The present article investigate this phenomenon: we show that in confined situations quantum-like eigenstates emerge from memory-mediated self-organisation.

**Results**

To achieve this confinement, we choose a case where the particle is trapped in an axisymmetric potential well. We obtain this situation by applying an attractive central force on a magnetized droplet. The characteristics of the walkers impose a specific approach for the investigation of their dynamics in a potential well. In the quantum situation a particle trapped in a harmonic potential well can be in successive eigenstates of increasing energy and decreasing mean wavelength as sketched in the 1D case in Fig. 1a. For a walker it is not possible to vary the kinetic energy of the particle: the droplet velocity is approximately constant and its wave field is characterized by a fixed wavelength $\lambda_F$. However it can be noted that in the quantum situation the successive eigenstates could be obtained at a fixed energy and mean wavelength by tuning the width of the well (see Fig. 1b). Here we will use this idea to seek the possible emergence of eigenstates when the spatial confinement is varied.

### Confinement of the droplet with a magnetic force

The basic experimental set-up is similar to that used in previous works[11,16-19]. A tank is filled with a silicon oil. It oscillates vertically at a frequency $f_0$ with an acceleration $\gamma = \gamma_m \sin(2\pi f_0 t)$. The amplitude $\gamma_m$ can be continuously tuned from a value of the order of the acceleration of gravity $g$ up to the Faraday instability threshold $\gamma_m^F$. The "walking" regime appears at a threshold $\gamma_m^W$ (located below $\gamma_m^F$), when the drop becomes a source of damped Faraday waves with periodicity $T_F$ and wavelength $\lambda_F$. Above $\gamma_m^W$, the drop velocity increases rapidly and saturates; the velocity modulus $V$ of a given droplet is approximately constant along its path (with 20% maximum variation), being imposed by the self-propulsion. An essential feature of the walker is the structure of its wave field. As demonstrated in ref. 11, Faraday waves centred at the impact point are created at each bounce. These waves damp out with a characteristic memory time $\tau$ related to the distance to the Faraday instability threshold. The parameter $M = \tau / T_F = \gamma_m / (\gamma_m^F - \gamma_m)$ indicates the number of past sources contributing to the global wave field. It is thus a measure of what we call the wave-mediated path memory of the particle.

In the present experiment, we want to submit the drop to a controlled force. For this purpose, the drop is loaded with a small amount of ferrofluid and becomes sensitive to any magnetic field. The encapsulation is obtained by dipping a conical needle into some ferrofluid immersed at the periphery of the bath. When the needle is swiftly pulled out, a liquid bridge containing a thread of ferrofluid is formed. It then breaks into a drop of oil containing a droplet of ferrofluid. The magnetic drop is completely covered by the silicon oil and the slight change in mean density and viscosity has no measurable effect on the dynamics of the bouncing. The presence of the ferrofluid has a negligible effect on the bouncing regimes and the drop velocity. The experimental cell is immersed in two superposed magnetic fields (Figs. 1c and 1d). Two large coils in Helmholtz configuration generate a vertical and spatially homogeneous field $\mathbf{B}_0$ perpendicular to the bath surface. A cylindrical magnet of diameter 15 mm, placed on the cell axis provides a second non-homogeneous magnetic field $\mathbf{B}_1(d, \mathbf{r})$, where $r$ is the distance to the axis and $d$ the tunable distance of the liquid surface. The characterization of the generated magnetic field is detailed in the Method section. The ferrofluid drop, polarized by the field, acquires a magnetic moment $\mathbf{m}_B$. Near the axis of the magnet, the resulting force can be approximated by a spring force:

$$\mathbf{F}_m(d) = -\kappa(d)\mathbf{r} \qquad (1)$$

The value of $\kappa$ depends on the volume $v_{ferro}$ of the encapsulated ferrofluid and its susceptibility. The force can be considered as harmonic up to a radius of approximately $3\lambda_F$. For a given drop, the spring constant can be tuned by changing the distance $d$ between the magnet and the oil surface. The angular frequency that characterizes the harmonic potential well is thus $\omega = \sqrt{\kappa(d)/m_W}$, where $m_W$ is the effective mass of the walker. Calibration is required as $v_{ferro}$ and $m_W$ cannot be measured directly. We have developed a technique that gives $v_{ferro}/m_W$ for each drop (see Methods for details).

**Influence of the memory on the walker's dynamics**

In short memory situations ($M < 10$), whenever the drop is released at the periphery of the cell, it converges into a circular orbit centred on the magnet axis, maintaining a constant speed $V$. The variation of the normalized radius $R/\lambda_F$ of the orbits as a function of the dimensionless width of the potential well $\Lambda = \left(V\sqrt{m_W/\kappa}\right)/\lambda_F$ can be deduced by a simple mechanical equilibrium where the centripetal acceleration equate the spring force (see Fig. 2a). It yields $R/\lambda_F = \Lambda$. The parameter $\Lambda$ plays the role of a control parameter for drops of various sizes and velocities. In addition, once established, this master curve can be used to obtain the calibration $\sqrt{m_W/\kappa}$ for any drop, regardless of its size, and its ferrofluid content (see Methods).

The memory is tuned by changing the amplitude of the vertical forcing. We first discuss the evolution of the low-memory circular orbits when increasing memory. As $M$ increases, the observed stable solutions become confined to narrowing ranges of radii, forming tongues separated by forbidden zones (see Fig. 2b, correspondingly the ranges of values of $\Lambda$ in which they are observed become narrower). The width of these tongues decreases with increasing memory, leading to a discrete set of orbits of radii $R_n$.

As the self-propelled drop becomes sensitive to its own past, the trajectories become more diverse. At long memory, the most generally observed trajectory, when no particular care is taken in tuning the potential well, has a complex aspect with a looped structure that will be discussed below. However in the medium memory range (30<$M$<100) simple trajectories such as those shown in Figure 3 (and Supplementary Movies 1, 2 and 3) are obtained, each existing in a narrow range of values of $\Lambda$. These orbits are either stable (Fig. 3a, 3c, 3e, 3g) or precess azimuthally (Fig. 3b, 3d, 3f, 3h). Note that only the smallest orbit, $n$=1, remains circular at long memory (Fig. 3a, 3b). The $n$=2 orbit assumes a stadium shape (Fig. 3c, 3d) while the next orbits are warped circles with symmetry 3, 4 etc…

The shapes of all the observed trajectories have a common feature: they can be fitted by generalized Cassini ovals[21]. These curves, originally introduced by Cassini, are the loci of the points where the product of the distances to two fixed foci is constant. The two foci being at a distance 2a from each other, their polar equation is:

$$r^4 + a^4 - 2a^2 r^2 \left[1 + 2\cos(2\theta)\right] = b^4 \qquad (2)$$

The ratio $a/b$ determines the shape of the curve. The circle is the trivial case where the two foci coincide. The stadium-shaped orbits are Cassini ovals (in Fig. 3c). The figure of eight shown in Fig. 3e is the Cassini curve with $a/b$=1 i.e. a Bernoulli lemniscate of polar equation $r^2 = 2a^2 \cos(2\theta)$. As for the warped circles and the trefoils, they can also be fitted by generalized Cassinian curves with three foci or more. The Cassinian ovals provide the best fits. However, other related families of curves having the same topology, such as the Cayley ovals, could also provide possible fits for the trajectories.

**A double quantization**

We characterize the particle's motion by two features, its mean non-dimensional distance to the central axis $\overline{R}$ :

$$\overline{R} = \frac{\sqrt{\langle R^2 \rangle}}{\lambda_{\mathrm{F}}} = \frac{1}{N}\sqrt{\sum_{k=1}^{N}\frac{r_k^2(t)}{\lambda_{\mathrm{F}}^2}}. \qquad (3)$$

and its mean non-dimensional angular momentum $\overline{L}_z$ :

$$\overline{L}_z = \frac{\langle L_z \rangle}{m_{\mathrm{W}}\lambda_{\mathrm{F}}V} = \frac{1}{N}\sum_{k=1}^{N}\left(\frac{\mathbf{r}_k}{\lambda_{\mathrm{F}}} \times \frac{\mathbf{V}_k}{V}\right). \qquad (4)$$

where $r_k$ is the position of the $k^{th}$ bounce and $N$ is the total number of bounces.

Similarly, the other types of stable trajectories undergo the same evolution with memory but there is a memory threshold for their appearance. The lemniscates (Fig. 3e, 3f) show-up for $M \geq 20$ in the range $0.4 < \Lambda < 0.7$. In wider potential wells, in the range of $1.1 < \Lambda < 1.5$, stable lemniscates of larger size are also visible, as well as another type of orbit in the shape of a trefoil (Fig. 3g, 3h). The range of $\Lambda$ in which these solutions are observed, becomes narrower when $M$ increases.

The remarkable feature of all these orbits is the double quantization of their spatial extent and angular momentum. Figure 4a plots the measured normalized radii $\overline{R}$ for the various trajectories as a function of the control parameter $\Lambda$ for intermediate memory ($30 < M < 70$). For orbits in the shape of a circle (or of warped circles), the mean radii satisfy $\overline{R}_n = (n - \varepsilon)\lambda_F/2$, with $n$ being the successive integers and $\varepsilon = 0.26 \pm 0.2$. The origin of this phenomenological relation is related to the Bessel functions as shown below. Note that the values of these radii are a function of the wavelength only: the control parameter $\Lambda$ simply selects amongst the possible $R_n$. Similarly, the other types of trajectories also correspond to discrete values of $\overline{R}$ (Fig. 4a). For instance, the mean radii of the eight-shaped paths are either $\overline{R} = 0.9 \pm 0.1$ or $\overline{R} = 1.9 \pm 0.2$, values that are close to the radii $R_n$ of the circular orbits $n=2$ and $n=4$, respectively. Trefoils are also quantized with $\overline{R} = 1.9 \pm 0.2$ corresponding to $n=4$ for the circular orbit.

In addition, we measure $\overline{L}_z$ for each type of orbit. The results are presented in Figure 4b. For the circular (or nearly circular) orbits, the mean angular momentum $\overline{L}_z$ of the drop is quantized with $\overline{L}_z = \pm(n - \varepsilon)/2$, the sign depending on the direction of rotation. For lemniscates, the symmetry of the trajectories induces $\overline{L}_z = 0$. For trefoils with $n=4$, we measure $\overline{L}_z = \pm(0.9 \pm 0.1)$, half the corresponding value for the nearly circular orbit with $n=4$. Note that the observed slow precession has only a minor effect on the value of $\overline{L}_z$.

A two dimensional graph can be obtained (Figure 4c) by plotting against each other the two main characteristics of all trajectories, i.e., their mean radius $\overline{R}$ and mean angular momentum $\overline{L}_z$. The various types of trajectories appear to be well defined in this 2D plot. For each level $\overline{R}_n$, only a discrete set of possible angular momentum values $\overline{L}_z$ is reachable by the system. Furthermore, the same discrete values of $\overline{L}_z$ are observed for different $\overline{R}_n$. From these results, we can define a second quantization number $m$ to order the angular momenta in a discrete set, where $m$ has the sign of $\overline{L}_z$. All the trajectories can be positioned at nodes $(n, m)$ of a lattice with the generic rules: $n$ is an integer, and $m$ can only take the values $m \in \{-n, -n+2..., n+2, n\}$ (see Fig. 4c). The orbits of the $n$=4 level have large perimeters. The very large values of $M$ necessary for their perfect quantization cannot be reached in our bath of finite size; hence the observed scatter of the results (Fig. 4c). The type of orbits corresponding to the $n$=3, $m$=1 mode is not observed as a stable mode presumably because its $\Lambda$ value overlaps with the ones of other simple modes. However a good candidate for this mode shows up in the decomposition of complex modes (see below). It takes the form of small loops of radius 0.37 $\lambda_F$ distributed at a distance 1.4 $\lambda_F$ from the centre. All the results given in Fig. 4 have been recovered in the simulations (see Supplementary Figure 1).

**Memory driving global wave field modes**

The model that permitted successful numerical simulations of the path memory effects in previous studies[11,17,19] is the starting point for both the numerical simulation and the understanding of these phenomena. In this approach, the horizontal and vertical motions of the droplet are decoupled. The vertical forcing of the bath determines each bouncing cycle, decomposed into a free flight and an interaction with the bath. During the free flight, the motion of the droplet is submitted to the magnetic force. At impact, the droplet generates a surface wave and has a dissipative sliding motion. The motion is first damped by friction and then modified by a kick on the slanted surface. The increase of the horizontal momentum is assumed to be proportional to the local slope at the point of impact[17,19]. This slope is determined by the global wave field, which is the sum of the elementary contributions induced by the previous collisions at times $t_j$ and positions $\mathbf{r}_j$, with $j$ ordering the impacts. Each collision induces a zero-order Bessel function $J_0$ at the Faraday wavelength, damped with a characteristic memory time $\tau$ and a characteristic spatial length $\delta$ resulting from viscosity. The global wave field at the position $\mathbf{r}_i$ of the i[th] impact time is

$$h(\mathbf{r}_i, t_i) = \sum_{j=-\infty}^{i-1} e^{-\frac{t_i - t_j}{\tau}} \, e^{-\frac{|\mathbf{r}_i - \mathbf{r}_j|}{\delta}} \, J_0\left(k_\mathrm{F}|\mathbf{r}_i - \mathbf{r}_j|\right), \qquad (5)$$

where $k_\mathrm{F} = 2\pi/\lambda_\mathrm{F}$ is the Faraday wave number. Such a model, which captures all of the required dynamical features of the walker behaviour, is supported by the more recent and complete hydrodynamic analysis of the bouncing[22,23].

The numerical simulation of this model provides results in excellent agreement with the experiments (see Supplementary Figure 1). However, this approach does not provide a direct insight on the selection of the observed eigenstates. We will now show that they result from a mean field effect. The droplet having a motion confined in a circular region of radius $\Lambda$ can excite the global wave eigenmodes of this domain. This will have a feedback effect on the droplet motion. The trajectories that eventually emerge are those for which the trajectory shape and the global wave field have achieved a mutual adaptation.

We will examine the quantization of the mean radii of the various types of trajectories in this framework. Here, we use a general approach that consists in decomposing the entire wave-field on the basis formed by the Bessel functions centred on the axis of the potential well. For the sake of simplicity, we consider the case of no spatial damping. In this case, the wave-field reduces to a sum of temporally damped Bessel functions $J_0$ emitted along the past trajectory. Using Graf's addition theorem[24], the global wave field $h(\mathbf{r}, t_i)$ at position $\mathbf{r} = (r, \theta)$ and time $t_i$ (eq. (5)) can be decomposed on the base formed by the Bessel functions centred on the axis of the potential well:

$$h(\mathbf{r}, t_i) = A_0 J_0(k_\mathrm{F} r) + \sum_{n=1}^{+\infty} J_n(k_\mathrm{F} r)\left[A_n \cos(n\theta) + B_n \sin(n\theta)\right] \qquad (6)$$

with the $A_n$ coefficients depending on the particular trajectory.

We first focus on the coefficient $A_0$. In the case of the smallest circular orbit, $A_0$ is proportional to $J_0$: $A_0 = J_0(k_\mathrm{F} R)/\left(e^{T_F/\tau} - 1\right)$ (see Fig. 5a). When $R$ coincides with the first zero of $J_0$, no mean wave is generated and the droplet is submitted to the central force only. For small deviations, the mode is excited, creating a radial wave profile $h_\mathrm{R}(r)$ and thus an effective potential $E_p(R) \propto J_0^2(k_\mathrm{F} R)$ (see Fig. 5b-5c). The additional force due to the wave field is centrifugal if the radius is too small and centripetal in the opposite case (see Fig. 5d). This gives an interpretation to the experimental results. The radius $R_1$ coincides with the first zero of $J_0$. The same reasoning can be applied for the other nearly circular orbits (2,2), (3,3) etc... Their mean radii coincide with the successive zeros of $J_0$ for which the phenomenological relation involving the parameter $\varepsilon$=0.26 is a good approximation.

We can also analyze the wave field associated with all the stable trajectories. For this purpose we use the experimentally recorded trajectories and reconstruct the associated wave field from the modal analysis in equation (6). We find that each type of trajectory is associated to a restricted set of global modes that depends on the symmetry of the trajectory. For instance, in the case of lemniscates, the two-fold symmetry, favours the even-order Bessel modes while for trefoils, the tri-fold symmetry dominates. Figures 5e and 5g show two experimentally observed trajectories, a circular orbit and a lemniscate respectively. These trajectories are superimposed on their reconstructed wave field. The spectral decompositions of these fields on centred Bessel modes are shown in Figs. 5f and 5h. Only a small number of modes is needed. In agreement with the symmetries, $J_4$ is the dominant mode for lemniscates and $J_6$ is dominant for trefoils. Only a small number of modes is needed for each type of orbiting motion. These modes pilot the droplet dynamics. It can be noticed that the interplay of the particle and the wave is responsible for the non-conservation of the instantaneous momentum. This effect is a direct consequence of the non-central nature of the force exerted by the waves. The fact that the averaged momentum is conserved means that a periodic exchange exists between the particle and the wave.

**Discussion**

In the previous experiments involving a Coriolis effect[19], the waves only produced an additional radial force and thus a discretization of the classical circular orbits. These orbits were the result of a simple radial force balance on the droplet. Here the eigenstates emerge from a dynamical self-organization process resulting from the interplay between the wavefield and the droplet motion. They are dually defined by a dominant global wavefield mode and a specific type of trajectory. In addition the quantization concerns mean characteristics of the mode and survive the jitter (Fig. 3). In addition, a similarity with quantum mechanics can again be noticed: the selection rules relating $m$ and $n$ are the same that link the energy and the angular momentum of a quantum particle confined in a 2D potential well[25].

Finally, we can examine the walker behaviour when $\Lambda$ is not tuned at any of the values corresponding to a pure eigenstate. In all these regions the trajectories have a very complex shape as shown in Fig. 6a (see Supplementary Movie 4). This trajectory is obtained for a value $\Lambda$ set between those of pure $n$=2 lemniscates and ovals. Direct observation reveals that regular motions still exist but only during limited time intervals. This can be analysed by using the temporal evolution of the angular momentum $L_z$ as shown in Fig. 6b for 250 s of trajectory 5a. The signal is not erratic, but composed of domains with $\overline{L_z} = \pm 1$ and others with a rapid oscillation of $L_z$ around zero. Extracted from the recorded data, the corresponding trajectory fragments (singled out in Fig. 6a) are ovals ($n$=2, $m$=±2) and lemniscates ($n$=2, $m$=0) respectively. The detuned trajectory is thus formed of a succession of sequences of pure eigenstates with intermittent transitions between them. Along the trajectory we measure both the mean spatial extension $\overline{R}$ and angular momentum $\overline{L_z}$ averaged over the typical orbiting time. As expected in the case shown in Fig. 6a, $\overline{R} \approx \overline{R_2}$. We also build, from the analysis of long recordings, the probability distribution function of the mean angular momentum $p(\overline{L_z})$. It exhibits well-defined maxima corresponding to the $\overline{L_z}$ values of the lemniscate and $n$=2 oval modes respectively. It evidences the decomposition in pure eigenstates resulting from the intermittency phenomenon. The same type of effect is observed for all tunings of $\Lambda$ with the possibility of having several modes involved in the decomposition. In this general case, the quantized self-organized eigenstates survive, forming the basis of a spontaneous probabilistic decomposition.

## Methods

### Experimental set-up

The basic experimental set-up is similar to that used in previous works[11,17-20]. A tank is filled with a silicon oil of viscosity $\mu$ =20 cP and surface tension $\sigma$=0.0209 N.m$^{-1}$. It oscillates vertically at a frequency $f_0$=80 Hz with an acceleration $\gamma = \gamma_m \sin(2\pi f_0 t)$. The amplitude $\gamma_m$ can be continuously tuned from a value of the order of the acceleration of gravity $g$ up to the Faraday instability threshold observed at $\gamma_m^F$ = 3.8 g. The Faraday wave has a periodicity $T_F = 2/f_0$=0.025 s and a wavelength $\lambda_F$=4.75 mm. The "walking" regime appears at a threshold $\gamma_m^W$ (located below $\gamma_m^F$), when the drop becomes a source of damped Faraday waves[11]. Above $\gamma_m^W$, the drop velocity increases rapidly and saturates; the velocity modulus $V$ of a given droplet is approximately constant along its path (with 20% maximum variation), being imposed by the self-propulsion. Here, we limit ourselves to droplets of velocity ranging from 7 to 13 mm.s$^{-1}$.

### Loading a droplet with ferrofluid

The central potential exerted on the droplet results from a magnetic force. A small quantity of ferrofluid is encapsulated in the droplet in order to exert a controlled magnetic force.

The ferrofluid is composed of a suspension of nanoparticles of an iron-cobalt alloy in a glycerol solution. The particles volume fraction is 0.22 and the density $\rho$=2.06. In order to obtain the encapsulation we proceed in the following way. A large drop of ferrofluid is deposited at the periphery of the experimental bath. Because of its larger density, it sinks and rests on the bottom of the cell. A conical needle is then dipped into the bath, its tip plunging in the ferrofluid. When the needle is swiftly pulled out, a liquid bridge forms, containing a thread of ferrofluid. It then breaks into a drop of oil containing a droplet of ferrofluid.

### Magnetic field characterization

The trap is created by two independent magnetic fields. A homogeneous $\mathbf{B}_0$ (with $B_0 \approx 50$ G) is created by two coils in the Helmholtz configuration. We have checked that a magnetically loaded walker moves rectilinearly in this field. An axisymmetric potential well is obtained when a small cylindrical magnet is placed above the fluid surface along the axis of the cell at a distance $d$ ($30 \leq d \leq 80$ mm). This magnet generates a second magnetic field $\mathbf{B}_1(d,r)$, of typically 20 G at a distance $d$=50 mm with a gradient of 0.1 G.mm$^{-1}$ at $r$=10 mm from the center. The ferrofluid drop polarized by the whole field $\mathbf{B}=\mathbf{B}_0+\mathbf{B}_1$, acquires a magnetic moment $m_B = v_{\text{ferro}} \chi_0 /(1+\chi_0)B/\mu_0$ where $v_{\text{ferro}}$ is the encapsulated volume of ferrofluid, $\chi_0$ is the susceptibility, and $\mu_0$ the magnetic permeability. The magnetic field is maximal on the symmetry axis and the drop is thus trapped by an attractive force:

$$F_m = -\left(\mathbf{m}_B.\nabla\right)\mathbf{B} = v_{\text{ferro}}\left(\frac{\chi_0}{\chi_0+1}\right)\frac{1}{\mu_0}\left(\mathbf{B}.\nabla\right)\mathbf{B}$$

where the typical value of $(\mathbf{B}.\nabla)\mathbf{B}$ required for confining a walker is 5 G$^2$mm$^{-1}$ for $d$=50 mm with $B_0$=50 G.

In order to check the potential profile, the two components of the magnetic field have been measured as a function of both $d$ and $r$. The measured fields have been fitted by the analytic expression of the field generated by a cylindrical magnetic source (see Supplementary Figure 2). The potential well can be considered as harmonic in a central region of radius 15 mm for this small magnet which correspond to three times the Faraday wavelength ($\lambda_F$=4.75mm). This approximation holds for larger distances with larger magnets. In the harmonic approximation, the spring constant $\kappa$ is given by:

$$\kappa(d) = -v_{\text{ferro}}\frac{\chi_0}{2(1+\chi_0)\mu_0}\left[\frac{d^2}{dr^2}\left(B_0 + B_1(d,r)\right)^2\right]_{r=0} = v_{\text{ferro}}\frac{\chi_0}{(1+\chi_0)\mu_0}\alpha$$

where $\alpha$ is the curvature at the vertex of the parabola of $-B^2$. Thus, the magnetic trap depends on the magnetic properties of the droplet $v_{\text{ferro}}\chi_0 /(\mu_0(1+\chi_0))$ and the magnetic field profile through $\alpha$. The latter can be determined and tuned by changing the distance $d$ of the magnet to the bath (see Supplementary Figure 2).

**Direct measurement of the magnetic force**

Although $v_{\text{ferro}}$ and $m_W$ are difficult to determine experimentally, the angular frequency $\omega = \sqrt{\kappa(d)/m_W}$ can be characterized accurately using a calibration technique. The magnetically loaded drop is submitted to a vertical excitation $\gamma_m$ below the walking threshold so that it is simply bouncing. In this regime the drop drifts towards the center of the potential well. The magnet located at a fixed distance $d$ from the bath is set in a horizontal oscillation at a low frequency $f_M$ with an amplitude $r_M$. As a result, the droplet oscillates at the same frequency $f_M$ with an amplitude $r_d$ and a phase shift $\varphi$ (see

Supplementary Figure 3).

The frequency response is fitted with a linear response of a forced harmonic oscillator submitted to viscous damping (solid lines). Thus, the natural frequency $\omega/2\pi$ of the droplet in such a potential is equal to the resonant frequency measured when the motion of the droplet and the magnet have a quadrature phase relationship. Since the mass of the droplet and the acting force do not depend on the acceleration of the bath, this result remains valid in the walking regime.

In addition, this method can be used to measure $\omega$ as a function of the distance $d$ and to confirm the relation between $\omega$ and the characteristics of the potential well $\alpha$ obtained from magnetic field cartography (see Supplementary Fig. 2). $\omega$ has a square-root dependency with $\alpha$ (see Supplementary Fig. 4).

**Calibration of the magnetic interaction**

This calibration was a preliminary before performing experiments with a walker. Using a drop previously calibrated, the bath acceleration is set to the typical value of the low memory regime ($M$=10). The drop starts to move and the motion converges quickly to a circular motion centered on the magnet axis, maintaining a constant speed $V$. The final trajectory is finally independent of the initial condition. The specific selection of these circular trajectories is directly related to the constraint that the wave-driven droplet moves with a constant velocity. Amongst all of the possible solutions of the classical oscillator, only circular paths satisfy this condition. For a walker in this regime, the radius $R$ is obtained by writing that the magnetic force provides the centripetal force:

$$\frac{R}{\lambda_F} = \frac{V}{\lambda_F} \sqrt{\frac{m_W}{\kappa(d)}} = \Lambda$$

Since $m_W/\kappa(d)$ has been obtained in the calibration process and $V$ is measured directly, there is no adjustable parameter. Figure 2a shows the normalized radius $R/\lambda_F$ of the orbits as a function of the dimensionless width of the potential well $\Lambda = \left(V\sqrt{m_W/\kappa}\right)/\lambda_F$, which plays the role of a control parameter for drops of various sizes and velocities. There is an excellent collapse of the data obtained with various drop sizes containing various quantities of ferrofluid. The linear relation between $R$ and $\Lambda$ confirms the harmonic feature of the external force. The simple mechanical equilibrium in which the centripetal acceleration is equal to the magnetic force would yield: $R/\lambda_F = \Lambda$

The expected proportionality is observed; however, the experimental radii are 10% larger than expected. This shift has already been observed in previous experiments involving a Coriolis force without calibration. The origin of this shift is discussed in Ref. 19. In practice, being established, this master curve can be used to obtain the calibration $\sqrt{m_{\mathrm{w}}/k}$ for any drop regardless of its size and the ferrofluid content. For this reason the rather tedious calibration does not have to be repeated for all drops. It is sufficient in order to obtain the resonance frequency $f_0$ of a drop, to measure directly at low memory the evolution of the orbit radius with $\kappa$.

## Acknowledgments:

We are grateful to J. C. Bacri, C. Wilhelm and F. Gazeau for sharing with us some of their knowledge on ferrofluids and to D. Charalampous, D. Courtiade, A. Lantheaume and L. Rhea for technical assistance. We thank the Chateaubriand Fellowship Program, AXA Research Fund, Institut Universitaire de France and the French National Research Agency for financial support (ANR FREEFLOW and LABEX WIFI).

## Author Contributions

All authors contributed equally to this work.

## Competing financial interests

Authors declare they have no competing financial interests

# Figures

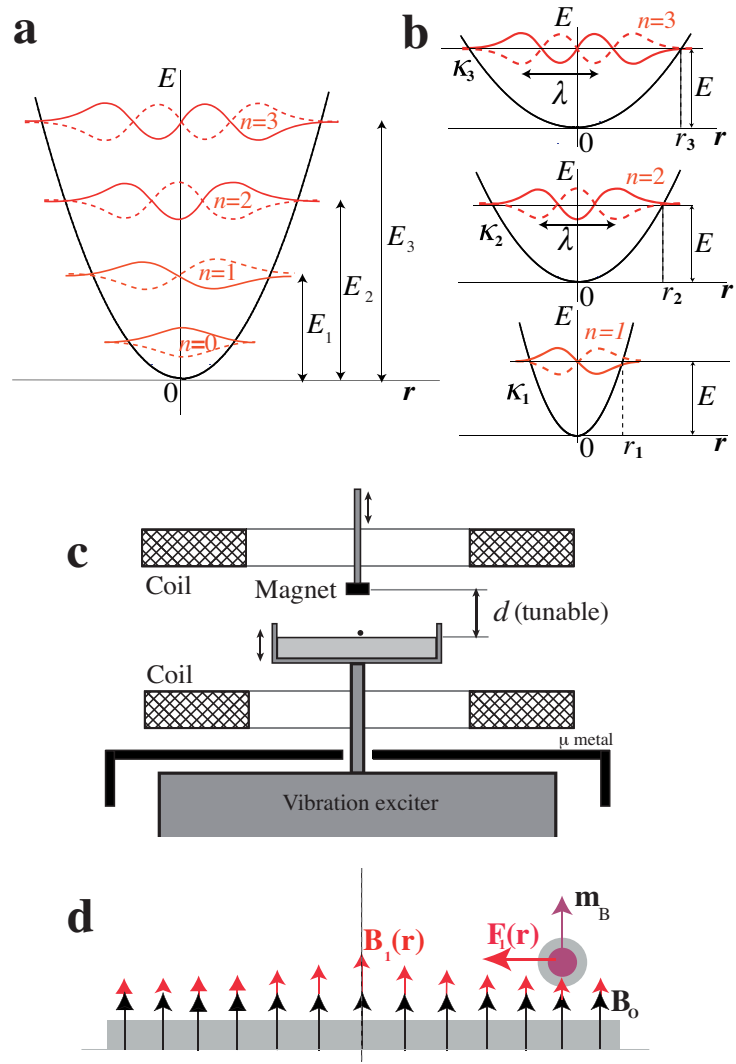

**Figure 1. Principle of the experiment and actual set-up. a**, Sketch of the successive eigenstates of increasing energy and decreasing wavelengths of a quantum particle in a 1D harmonic potential well. **b**, These successive eigenstates could be observed at a fixed energy and mean wavelength if the width of the potential well is varied. This experiment relies on the possibility to trap a walker in a harmonic potential well of tunable width. Since the wave field associated to a walker has a fixed wavelength $\lambda_F$ it is possible to expect that there exists a discrete set of potential wells characterized by their spring constants $\kappa_n$, for which discrete eigenmodes appear. **c,** Sketch of the experimental set-up. The Helmholtz coils generate a spatially homogeneous magnetic field $\mathbf{B}_0$. A magnet located at an adjustable distance $d$ from the bath surface generates a spatially inhomogeneous magnetic field $\mathbf{B}_1$. **d,** A central magnetic force is exerted on the droplet filled with a ferromagnetic fluid, induced by the superposition of a constant vertical magnetic field $\mathbf{B}_0$ and $\mathbf{B}_1$ It has, to a good approximation, a harmonic profile.

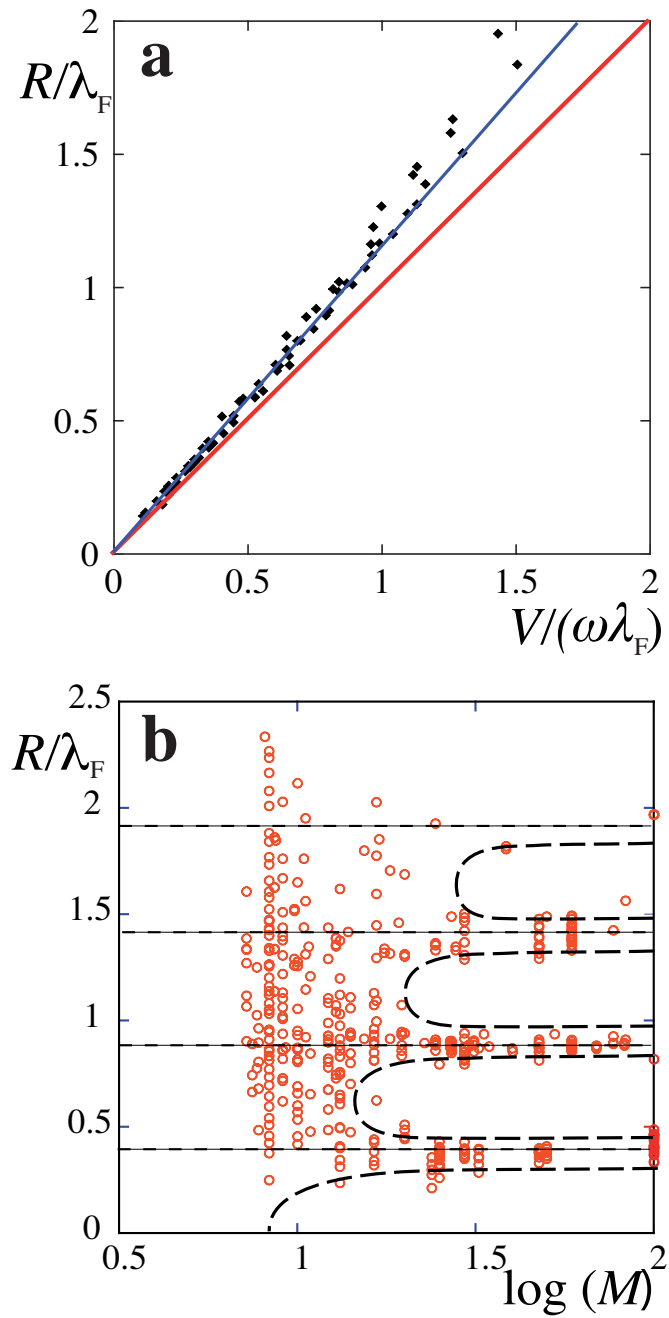

**Figure 2. Discretization of circular orbits with memory. a**, Plot of the dimensionless radius $R/\lambda_F$ as a function of the non-dimensional width of the potential well $\Lambda = V/(\omega\lambda_F)$ in the low-memory situation ($M=10$): experimental data (dots). The dependence follows the expected proportionality of a classical particle $R/\lambda_F = \Lambda$, with a slight shift of 10%. Inset: a typical circular orbit, with axis in $\lambda_F$ unit. **b**, Diagram illustrating the progressive discretization of the radii of the circular orbits as the memory parameter $M$ is increased. The experimental (circles) results were obtained for a range of the control parameter. The dashed lines provide guides for the eye.

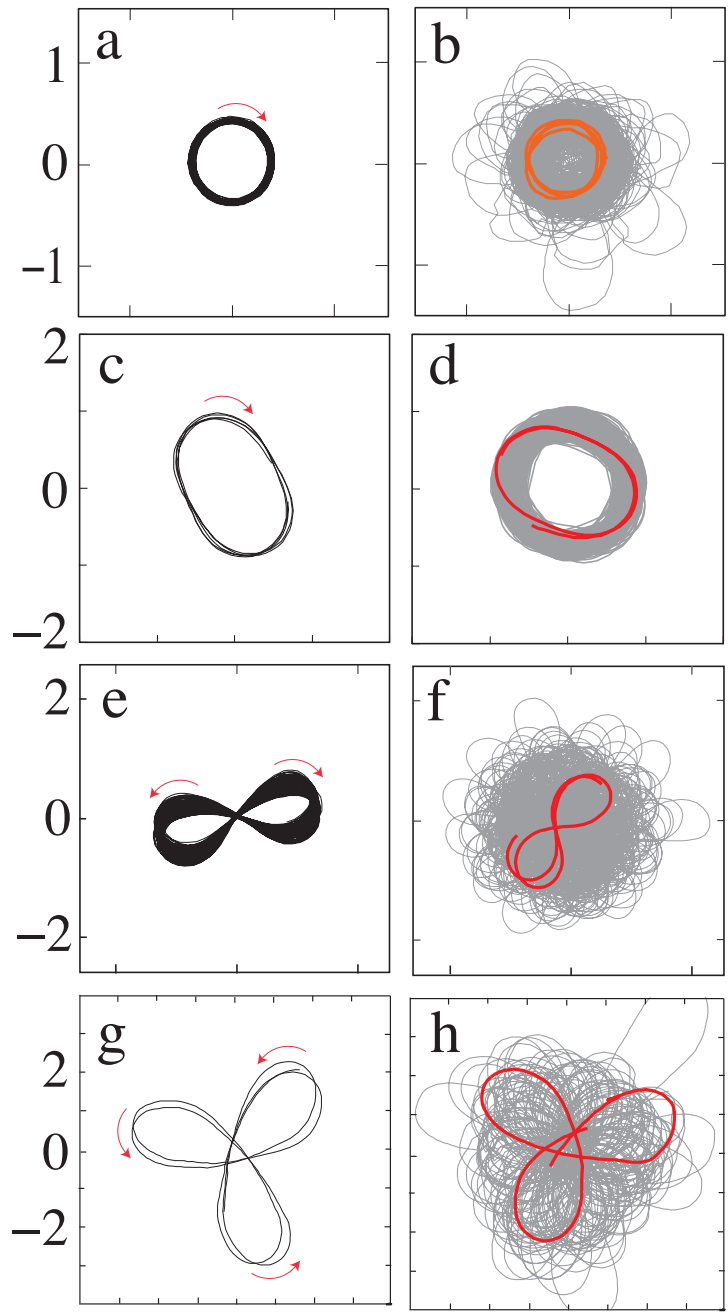

**Figure 3. Four of the experimentally observed long memory stable orbits.** Simple trajectories can be observed in narrow ranges of values of the tuning parameter $\Lambda$. They are shown in their most stable form (for $M \approx 50$) and in situations where they are slightly unstable (for $M \approx 100$). The scales are in units of the Faraday wavelength $\lambda_{\mathrm{F}}$=4.75 mm. **a, b,** The smallest circular orbit $(n=1, m=\pm 1)$. **c,d,** the oval orbit $(n=2, m=\pm 2)$ well fitted by a Cassini oval, **e,f,** The small lemniscates $(n=2, m=0)$, **g, h,** The trefoil $(n=4, m=\pm 2)$. In **b, d, f, and h** the trajectories are shown in grey for a long time interval and a part of it has been singled-out in red. The scales are in units of $\lambda_{\mathrm{F}}$. See Supplementary Movies 1-3 associated with trajectories of Fig. **a**, **c-d** and **f** respectively.

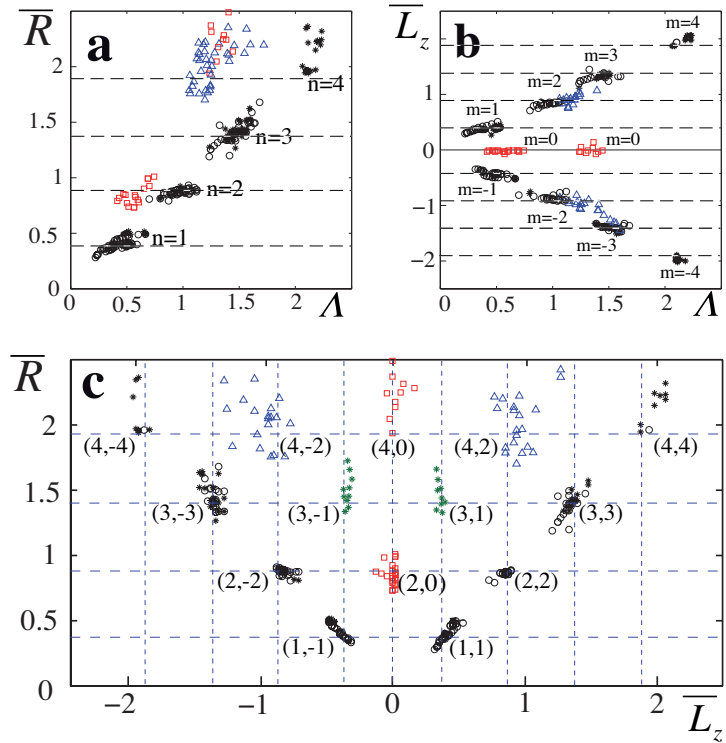

**Figure 4. The double quantization at long memory. a,** Evolution of the mean spatial extent $\overline{R}$ as a function of the control parameter $\Lambda$. The black circles correspond to the circular or oval orbits, the red squares represent the lemniscates and the blue triangles represent the trefoils. The stars are radii measured in organized sections of disordered patterns of the type shown below in Fig. 5. **b,** Evolution of the mean angular momentum $\overline{L}_z$ as a function of $\Lambda$ for the same set of trajectories. **c,** The same set of trajectories is now characterized by plotting their angular momentum $\overline{L}_z$ as a function of their mean spatial extent $\overline{R}$. The symbols are the same as those in **a**. The trajectories are labelled by their position $(n, m)$ on the lattice. The type of orbits corresponding to the ($n$=3, $m$=1) mode is not observed as a stable mode. A good candidate (green stars) for this mode shows up in the decomposition of complex modes and takes the form of small loops of radius 0.37 $\lambda_F$ distributed at a distance 1.4 $\lambda_F$ from the centre. The simulations using path-memory model show a similar results (see Supplementary Figure 1).

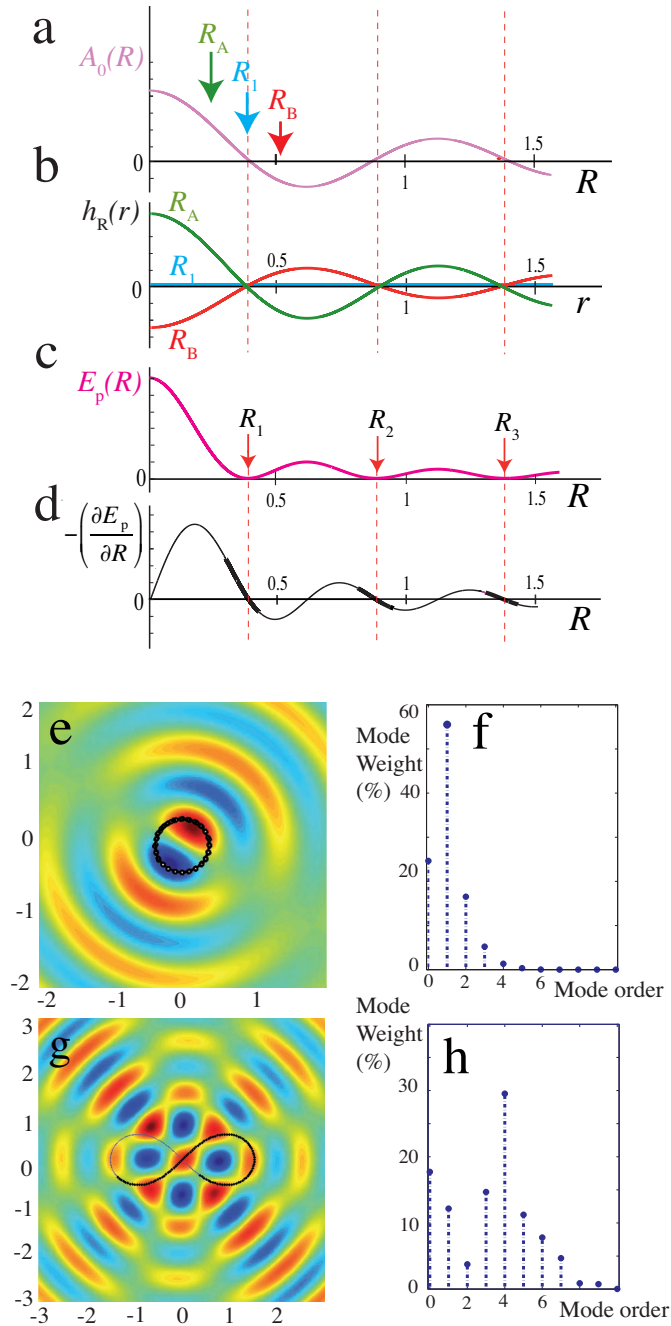

**Figure 5. Mode analysis of the wave field. a-d,** Discretization of the radius of the smallest circular orbit (1,1). The radial force is dominated by the centred zero-order Bessel $J_0$ function. The generated global wave field is $h_0(r)=A_0(R)J_0(k_F r)$. **a,** the amplitude $A_0(R)$ depends on the radius $R$ of the trajectory: $A_0(R) \propto J_0(k_F R)$. In **b** are shown three surface profiles $h_R(r)$ created by drops orbiting at three radii: $R_1$ that correspond to the first zero of $J_0(k_F r)$ and $R_A$ and $R_B$ that are slightly smaller and slightly larger respectively (see **a**). Note that the circular orbit of radius $R_1$ does not excite the $J_0$ mode of the global wave field. **c,** whenever R does not coincide with a zero of the $J_0$ mode, a mean wave field is generated so that the drop has an additional potential

energy $E_\mathrm{p}(R) \propto J_0{}^2(k_\mathrm{F}R)$. **d,** If the radius is slightly smaller or larger than one of the radii $R_n$, the $J_0$ mode is excited and exerts an additional "quantization" force onto the droplet $: -(\partial E_\mathrm{p}/\partial R) \propto J_1(k_\mathrm{F}R)J_0(k_\mathrm{F}R)$. **e,** the experimental trajectory of an experimental circular orbit $\left(n=1, m=1\right)$ at a memory parameter $M$=32. Scales are in $\lambda_\mathrm{F}$ units. **f,** spectral decomposition of the wave-field in centred Bessel functions. The trajectory being close to the ideal, the amplitude of the $J_0$ mode is weak, the dominant $J_1$ mode is responsible for the azimuthal propulsion of the droplet. **g,** An experimentally observed lemniscate trajectory $\left(n=2, m=0\right)$. The latest $M$ impacts are shown as open dots. It is superimposed on the reconstructed global wave-field. **h,** the spectral decomposition of this wave field showing that for this near ideal orbit the $J_2$ mode is specifically weak.

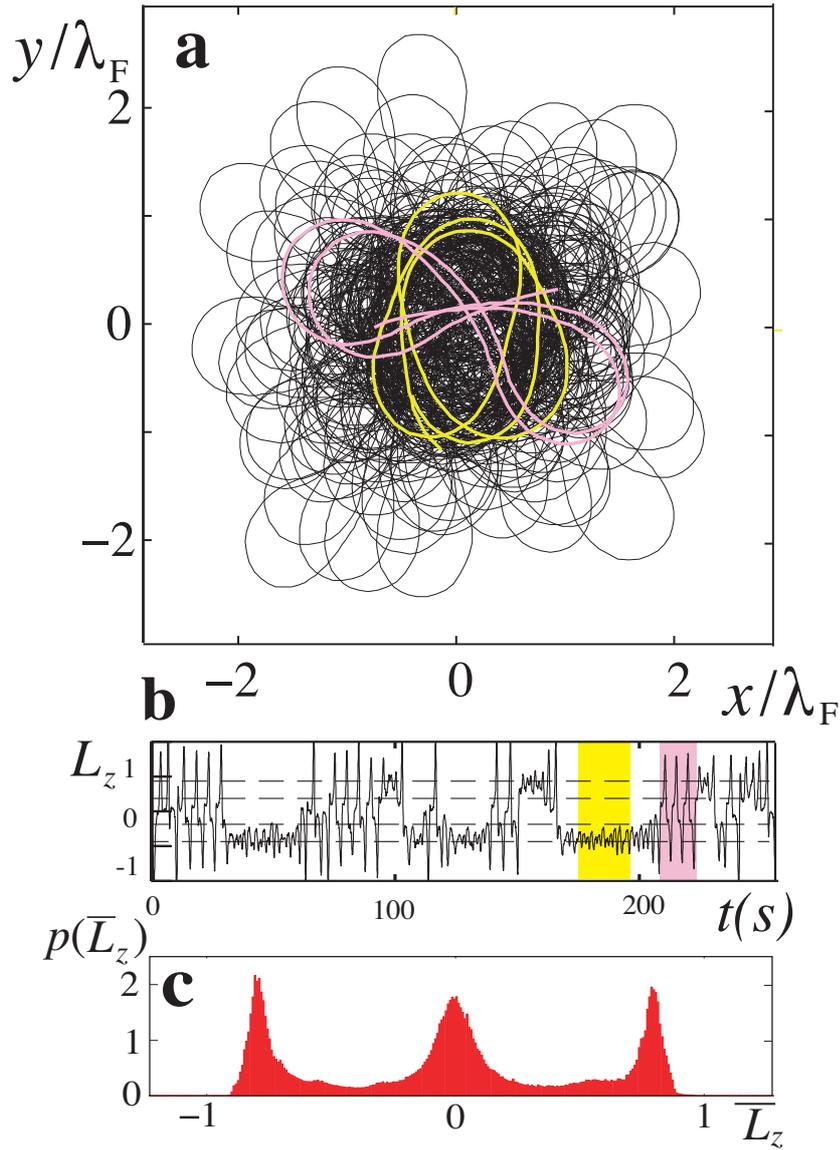

**Figure 6. Eigenstates decomposition of a complex trajectory. a,** The complex aspect of the type of trajectory most generally observed at long memory recorded during 800 seconds (see Supplementary Movie 4). **b,** A part of the corresponding time recording of the angular momentum $L_z$ showing the intermittent transitions between two types of orbiting motion. The time intervals in yellow and pink correspond to the oval and lemniscate-shaped trajectory segments singled out in (a). **c,** The probability distribution of the average of the mean angular momentum $\overline{L}_z(\Delta t)$ measured when a slot of length $\Delta t = 4\pi\overline{R}_2/V$ is slid along the trajectory. The maxima correspond to the ovals ($n$=2, $m$=±2) and lemniscate ($n$=2, $m$=0) modes respectively. As in the case of circular orbits, the mean radius of the oval is equal to the second zero of the $J_0$ Bessel function and their associated non dimensional angular momentum can be approximated by $\overline{L}_z = \pm(1 - \varepsilon/2)$ with $\varepsilon$=0.26.